\documentclass[a4paper,11pt]{article}
\usepackage{jinstpub} 
\usepackage{subcaption}
\usepackage[utf8]{inputenc}
\usepackage{upgreek}
\notoc
\usepackage{graphicx}

\title{Exploring unique design features of the Monolithic Stitched Sensor with Timing (MOST): yield, powering, timing, and sensor reverse bias}
\collaboration[c]{on behalf of the ALICE collaboration}
\author[a,c]{Mariia Selina}
\author[a,c]{R. Barthel}
\author[b]{S. Bugiel}
\author[b]{L. Cecconi$^{1}$}
\author[b]{J. De Melo$^{2}$}
\author[a]{M. Fransen}
\author[a,c]{A. Grelli}
\author[a,d]{I. Hobus}
\author[a]{A. Isakov}
\author[b]{A. Junique}
\author[b]{P. Leitao}
\author[b]{M. Mager}
\author[b]{Y. Otarid}
\author[b]{F. Piro$^{3}$}
\author[a,c]{M.J. Rossewij}
\author[a,d]{S. Solokhin}
\author[a,d]{J. Sonneveld}
\author[b]{W. Snoeys}
\author[b]{N. Tiltmann}
\author[a]{A. Vitkovskiy}
\author[a,d]{H. Wennloef}

\affiliation[a]{Nikhef, Amsterdam, Netherlands}
\affiliation[b]{CERN, Geneva, Switzerland}
\affiliation[c]{Utrecht University, Utrecht, Netherlands}
\affiliation[d]{University of Amsterdam, Amsterdam, Netherlands}
\emailAdd{mariia.selina@cern.ch}
\abstract{
Monolithic stitched CMOS sensors are explored for the upgrade of Inner Tracking System of the ALICE experiment (ITS3) and the R\&D of the CERN Experimental Physics Department. To learn about stitching, two 26\,cm long stitched sensors, the Monolithic Stitched Sensor (MOSS), and the Monolithic Stitched Sensor with Timing (MOST), were implemented in the Engineering Round 1 (ER1) in the TPSCo 65nm ISC technology. Contrary to the MOSS, powered by 20 distinct power domains accessible from separate pads, the MOST has one global analog and digital power domain to or from which small fractions of the matrix can be connected or disconnected by conservatively designed power switches to prevent shorts or defects from affecting the full chip. Instead of the synchronous readout in the MOSS, the MOST immediately transfers hit information upon a hit, preserving timing information. The sensor reverse bias is also applied through the bias of the front-end rather than by a reverse substrate bias. This paper presents the first characterization results of the MOST, with the focus on its specific characteristics, including yield analysis, precise timing measurements, and the potential of its alternative biasing approach for improved sensor performance.
}
\begin{document}
\maketitle
\vspace*{\fill}
{\small
\noindent
${}^{1}$ Now with University of Geneva, Switzerland. \\
${}^{2}$ Now with Brookhaven National Laboratory, US. \\
${}^{3}$ Now with Miromico IC, Zurich, Switzerland.
}
\flushbottom

\section{Introduction}
During LS3 (2026–2030), the ALICE Inner Tracking System (ITS) will undergo a major upgrade with the replacement of its three innermost layers as part of the ITS3 project \cite{its3tdr}. This upgrade aims to enhance vertexing precision, increase physics yield, and improve overall detector performance in preparation for the High Luminosity LHC (HL-LHC) .

The ITS3 project aims to position the three inner layers of the detector closer to the interaction point, with the first layer at 19 mm, while reducing the material budget to 0.09\% $X_0$ per layer. This is achieved by employing a wafer-scale, stitched monolithic sensor chip that integrates all electrical functions directly on-chip. The chip is thinned to 50\, $\upmu$m and bent around the beam pipe, creating a vertex detector composed of cylindrical layers formed almost entirely by the silicon sensors themselves.

The TPSCo 65 nm CMOS imaging technology has been qualified for high-energy physics \cite{apts_paper,AGLIERIRINELLA2023168589,SNOEYS201790}
in the framework of the R\&D of the CERN Experimental Physics department, and the ALICE ITS3 upgrade project. Engineering Round 1 (ER1) includes two wafer-scale, 26\,cm long, stitched sensors, the Monolithic Stitched Sensor (MOSS) \cite{leitao_development_2024,terlizzi_characterization_2024}
and the Monolithic Stitched Sensor with Timing (MOST). The MOST contains 901120 pixels distributed across 10 stitched Repeated Sensor Units (RSUs) of 25.5\,mm by 2.5\,mm each, designed with a pixel pitch of 18\,$\upmu$m $\times$ 18\,$\upmu$m. The MOST design explores a highly granular powering scheme to address yield, 
an asynchronous readout scheme offering timing capabilities, and an alternative approach to reverse sensor biasing. This contribution describes the initial measurement results of the MOST, focusing on these specific characteristics, to present an outlook for future advancements in CMOS monolithic sensors.

\section{Highly granular powering scheme using power switches and dense matrix circuitry}
Contrary to the MOSS, segmented in 20 fully separated power domains accessible by separate pads, the MOST is equipped with a much more granular powering scheme: power switches, fully programmable through the slow control interface of the chip, can connect or disconnect small portions of the matrix
to or from the global, conservatively designed power network.
The aim is to isolate shorts or defects and prevent them from affecting the full chip, allowing the internal circuitry powered through the power switches to be designed at the full density, so at the maximum density defined by the design rules. 

\begin{figure}[htbp]
\centering
\includegraphics[width=0.8\textwidth]{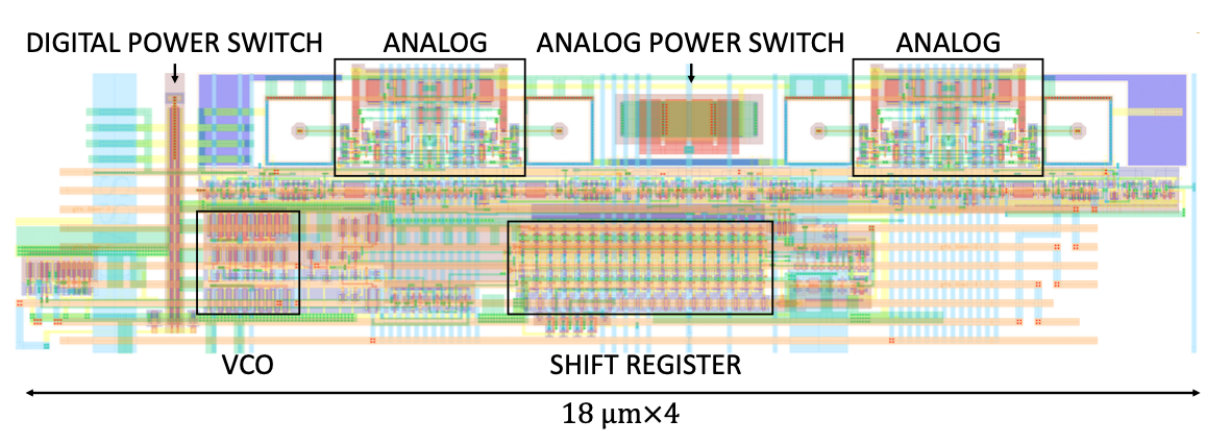}
\caption{MOST Layout of a group of 4 pixels \cite{piro_front-end_nodate}.}
\label{fig:pixel_layout}
\end{figure}

Figure~\ref{fig:pixel_layout} shows the layout of a group of 4 pixels, and illustrates the conservative design of the power switches, and the dense design of the rest of the circuitry. 
\begin{figure}[htbp]
\centering
\begin{subfigure}[b]{0.45\textwidth}
    \centering
    \includegraphics[width=\textwidth]{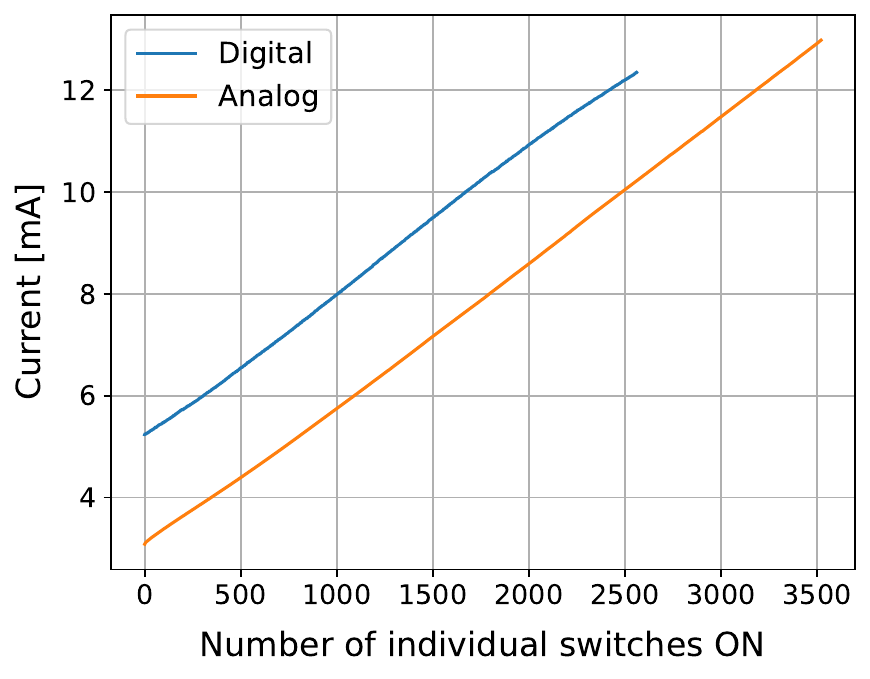}
    \end{subfigure}
    \hfill
        \begin{subfigure}[b]{0.50\textwidth}
        \includegraphics[width=\textwidth]{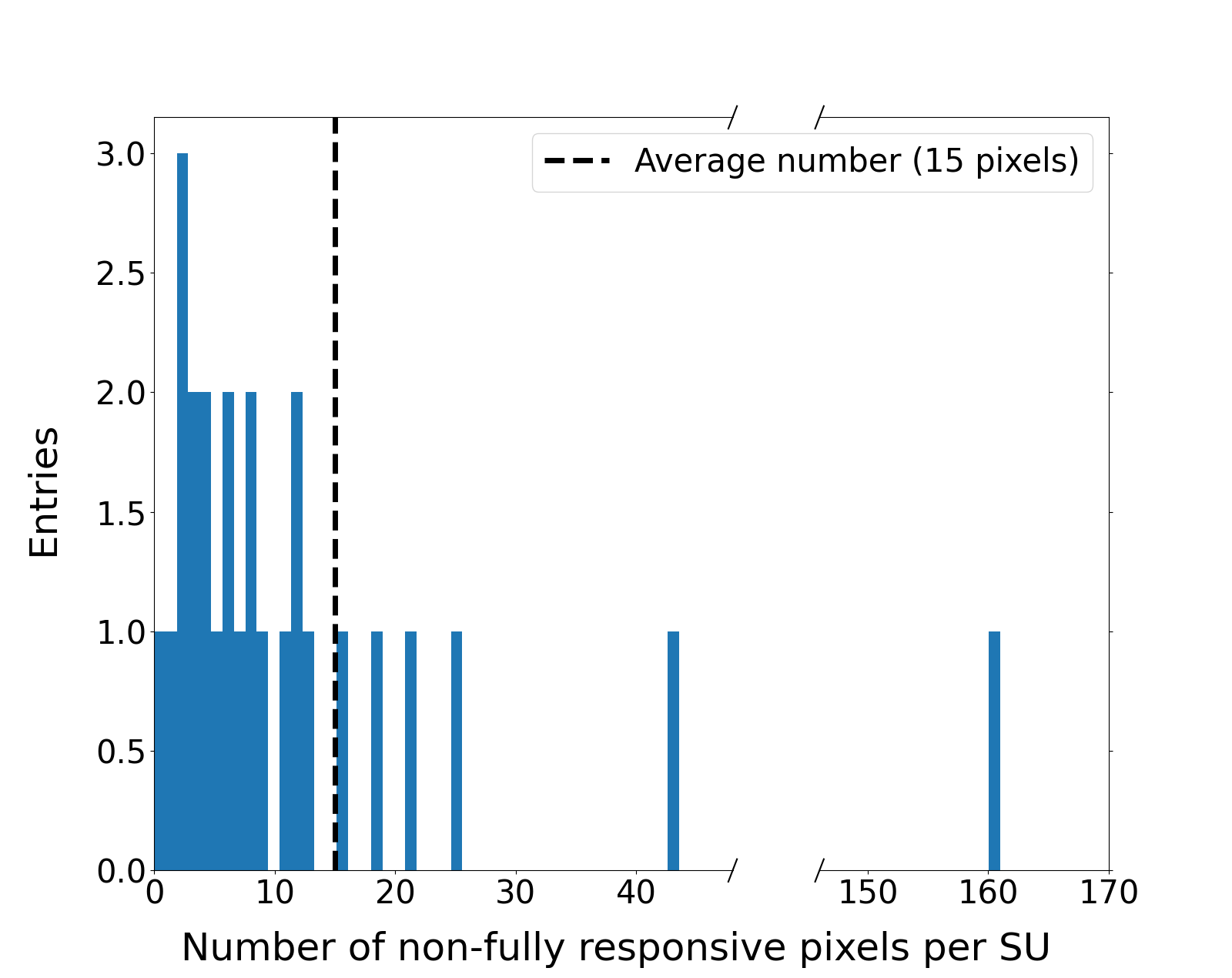}
    \end{subfigure}
    \caption{Graph on the left showing the increase in power consumption of the chip with the number of pixels powered on. On the right there is first evaluation of the individual pixel yield for 4 chips or 40 stitched units (SU): the average number of non-fully responsive pixels per stitched unit is 15 per 90112 pixels, indicating a pixel yield better than 99.98\%.}
    \label{fig:power_nonresp}
\end{figure}

The digital circuitry of the MOST is power gated along the pixel columns in subsets of 352 pixels, resulting in a total of 2,560 independently controlled digital switches. The analog power is grouped per 256 pixels, arranged in 4 rows of 64 pixels, amounting to 3,520 analog switches. This difference in segmentation of analog and digital power resulted in some design complication, and has been avoided in the MOSAIX, the full-size, fully functional prototype for the ITS3 upgrade, which adopts the same principle of power switches and global power supplies with reduced granularity.

To evaluate the validity of this powering approach, power-on tests were performed first.

A processing issue, expected to be resolved in the next fabrication, caused shorts on the global power network for approximately 30\% of the chips measured so far. This issue also affected the MOSS and was previously reported in Ref. \cite{eberwein_yield_2024}. The 30\% failure rate is consistent with observations on the MOST chip, where a significant overlap in the global power network between two metals prone to shorts has been identified. 

All MOSTs which could be powered on with all power switches in the matrix turned off, were found to be functional, and allowed  gradual powering of the full chip using the power switches.

The power consumption measurements presented in Figure~\ref{fig:power_nonresp} (left) exhibit linear increase in power consumption as more areas are activated by the power switches, and confirms the effectiveness of this fine-granularity power distribution approach in managing power.

As a further validation, particularly regarding the feasibility of designing the pixel circuitry at the full density, the detailed pixel yield was evaluated on four different chips.
Figure~\ref{fig:power_nonresp} (right) shows the number of non-responsive pixels per stitched unit or per 90112 pixels for these 4 chips, i.e. for 40 stitched units. The average number is 15 pixels per 90112 indicating a pixel yield above 99.98\%. The 160 non-fully responsive pixels forming an outlier stitched unit, are actually within the same 3 or 4 rows, which hints at a defect in one of the shared lines within those rows. This one outlier increases the average non-fully responsive pixels per RSU from 11 to 15. 
While this yield analysis is done with limited statistics, the excellent individual pixel yield demonstrates the feasibility of high-density designs for future applications. 

\section{Timing Capabilities with Asynchronous Readout}
\begin{figure}[htbp]
    \centering
    \begin{subfigure}[b]{0.25\textwidth}
        \centering
        \includegraphics[width=\textwidth]{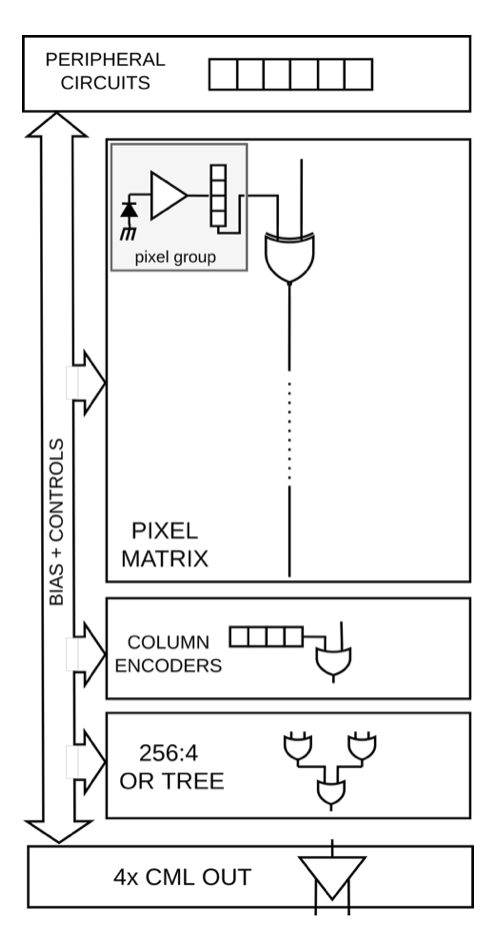}
        \label{fig:most_readout_logic}
    \end{subfigure}
    \begin{subfigure}[b]{0.63\textwidth}
        \centering
        \includegraphics[width=\textwidth]{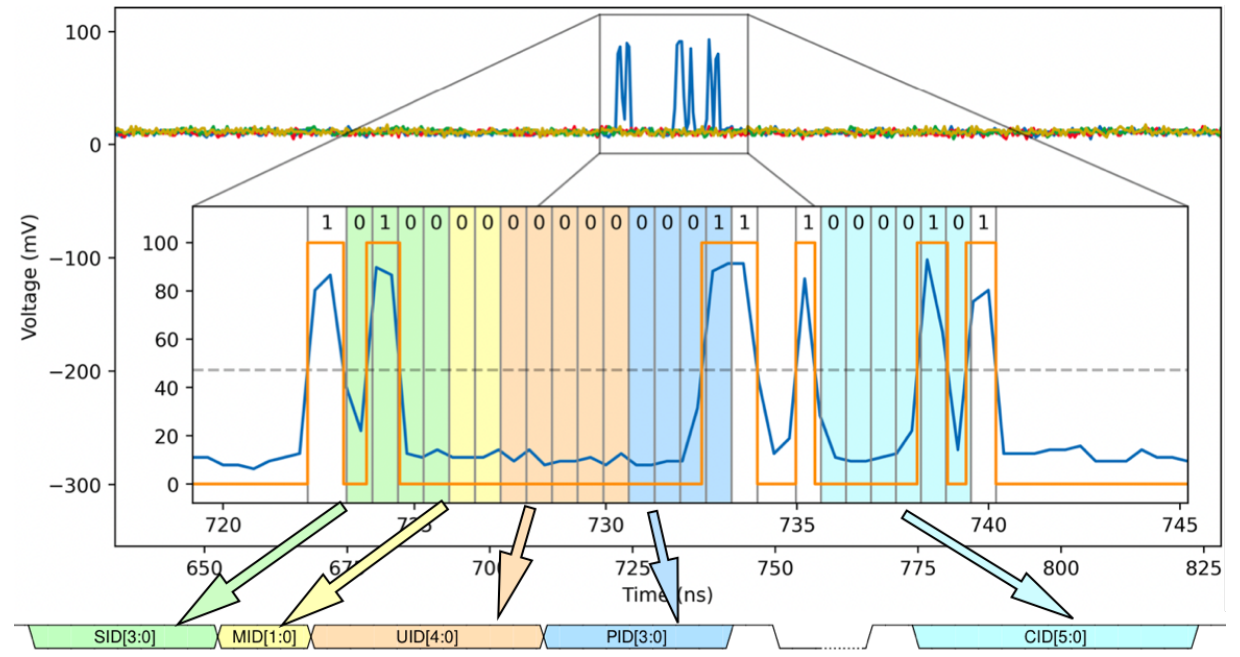}
        \label{fig:signal_waveform}
    \end{subfigure}
    \caption{Figure on the left shows the schematic of signal propagation from the pixel to the chip output. 
    Figure on the right illustrates how hit information is encoded within the output waveform, including the address of the RSU (SID), Matrix ID (MID), Unit ID (UID), Pixel ID (PID), and Column ID (CID).}
    \label{fig:most_combined}
\end{figure}

The asynchronous readout in the MOST is designed to provide precise timing information by transmitting a bit sequence with the pixel row address, immediately upon detection of a hit. The same sequence can also be transmitted after the in-pixel comparator returns to zero, providing time-over-threshold information. This data is transmitted through a column, at a distance of up to $\sim$26\,cm, to the endcap, where the column address is added to the bit sequence and where all hits go through an OR tree (logic) into one of the 4 CML drivers, transmitting the data off-chip as shown in Figure~\ref{fig:most_combined} (left). The MOST chip has four independent transmission lines per column to minimize collisions from hits on neighboring pixels. Figure~\ref{fig:most_combined} (right) illustrates how the information is encoded in the output waveforms. Although the asynchronous readout will not be implemented in the ITS3 chip, it is very relevant for future developments to understand to what level the timing information is preserved by the on-chip transmission over such a large distance.

\begin{figure}[htbp]
    \centering
    \begin{subfigure}{0.52\textwidth}
        \centering        \includegraphics[width=\textwidth]{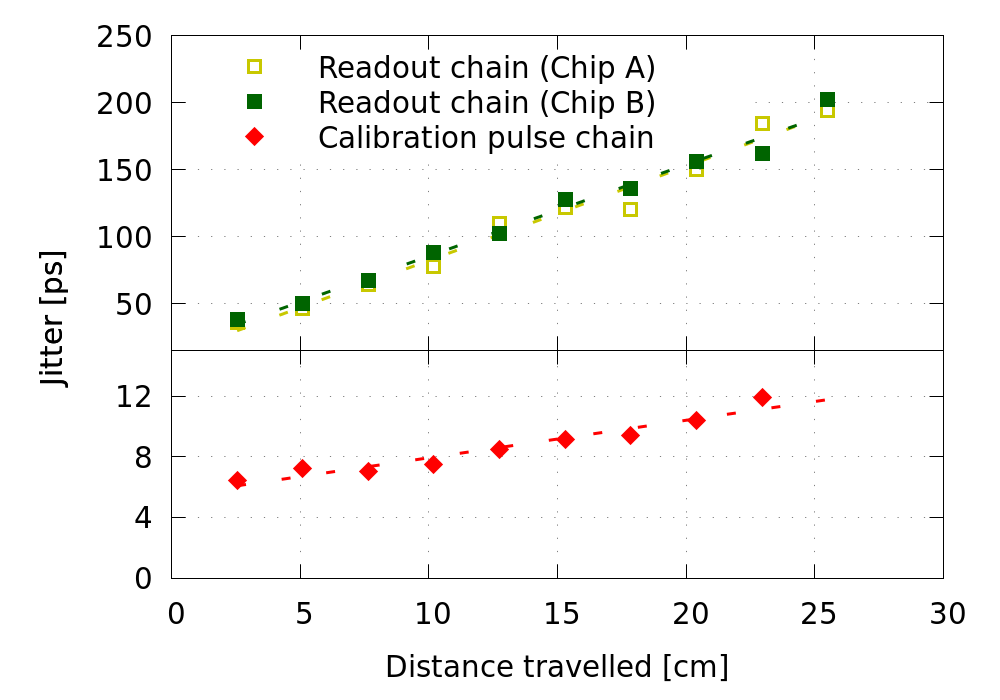}

        \label{fig:jitter}
    \end{subfigure}%
    \hfill
    \begin{subfigure}{0.48\textwidth}
        \centering
        \includegraphics[width=\textwidth]{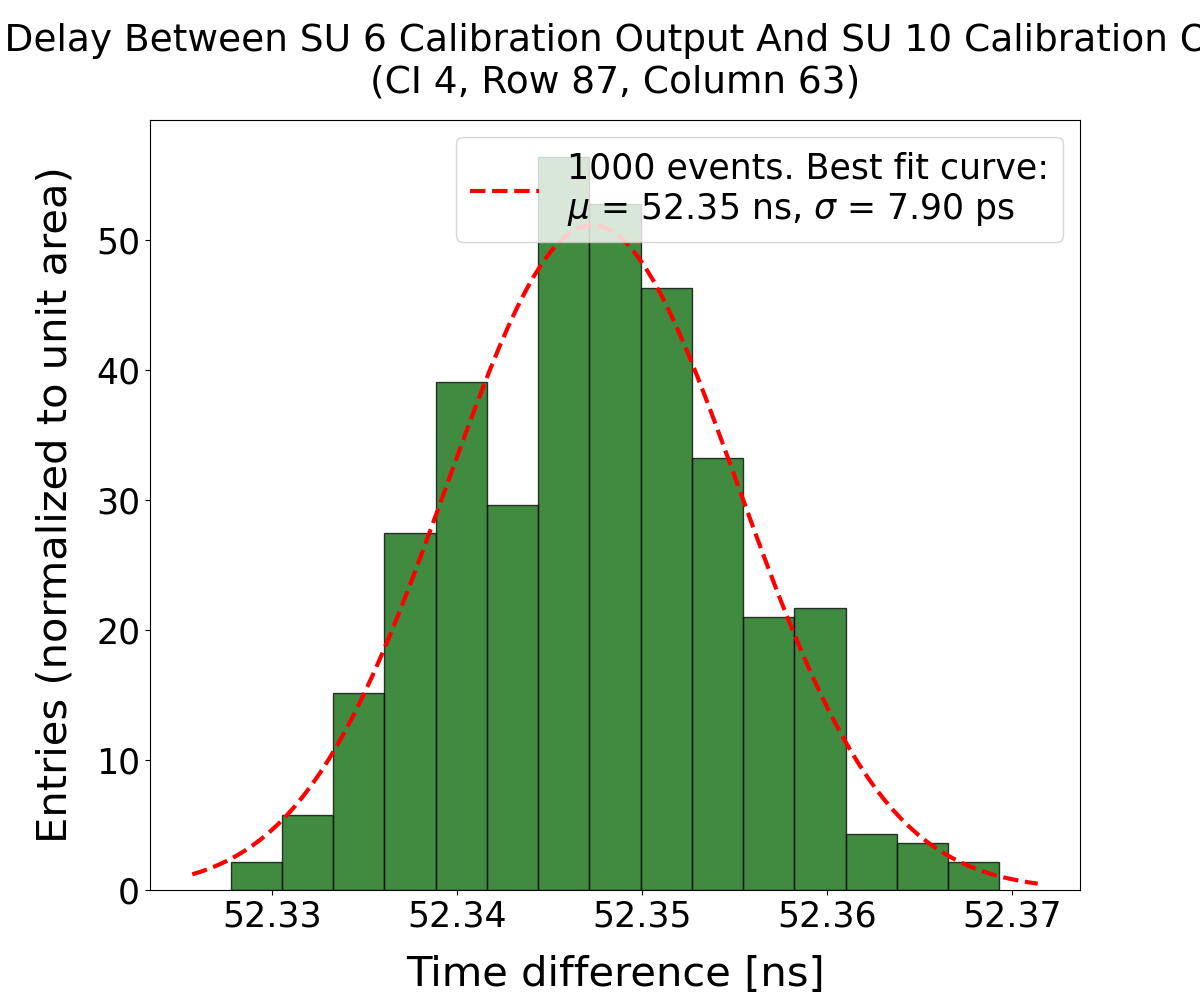}
        \label{fig:jitter_hist}
    \end{subfigure}
    \caption{First results on the MOST timing: low jitter is observed in the calibration pulse chain (right), at least an order of magnitude lower than the jitter of signals in the readout chain (left).} 
    \label{fig:jitters}
\end{figure}

 In the MOST sensor, except of the readout chain, there is also a CMOS calibration pulse chain that spans over the entire device.  The timing precision of both was evaluated by measuring the time delay between input and output signals. Figure~\ref{fig:jitters} (left) shows the preliminary jitter results of the readout chains on two different devices together with the performance of the calibration chain, while in Fig.~\ref{fig:jitters} (right) there is an example of the time difference distribution out of which the jitters have been extracted. There is a difference by more than an order of magnitude in the jitter of the calibration signal, which at the full distance of $\sim$26\,cm reaches 12\,ps, compared to that of the output signals in the readout chain which is about 20 times larger. 
 This is currently not understood and is under further study. These results demonstrate that the asynchronous readout is fully operational, but further characterization is required to understand its timing performance. Despite this, the  $\sim$12\,ps jitter observed on the calibration pulse over a distance of 26\,cm is encouraging for the long-distance on-chip transmission of signals with timing information. 

\section{Alternative Sensor Biasing Scheme}

\begin{figure}[htbp]
    \centering
    \begin{subfigure}[b]{0.50\textwidth}
        \centering
        \includegraphics[width=\textwidth]{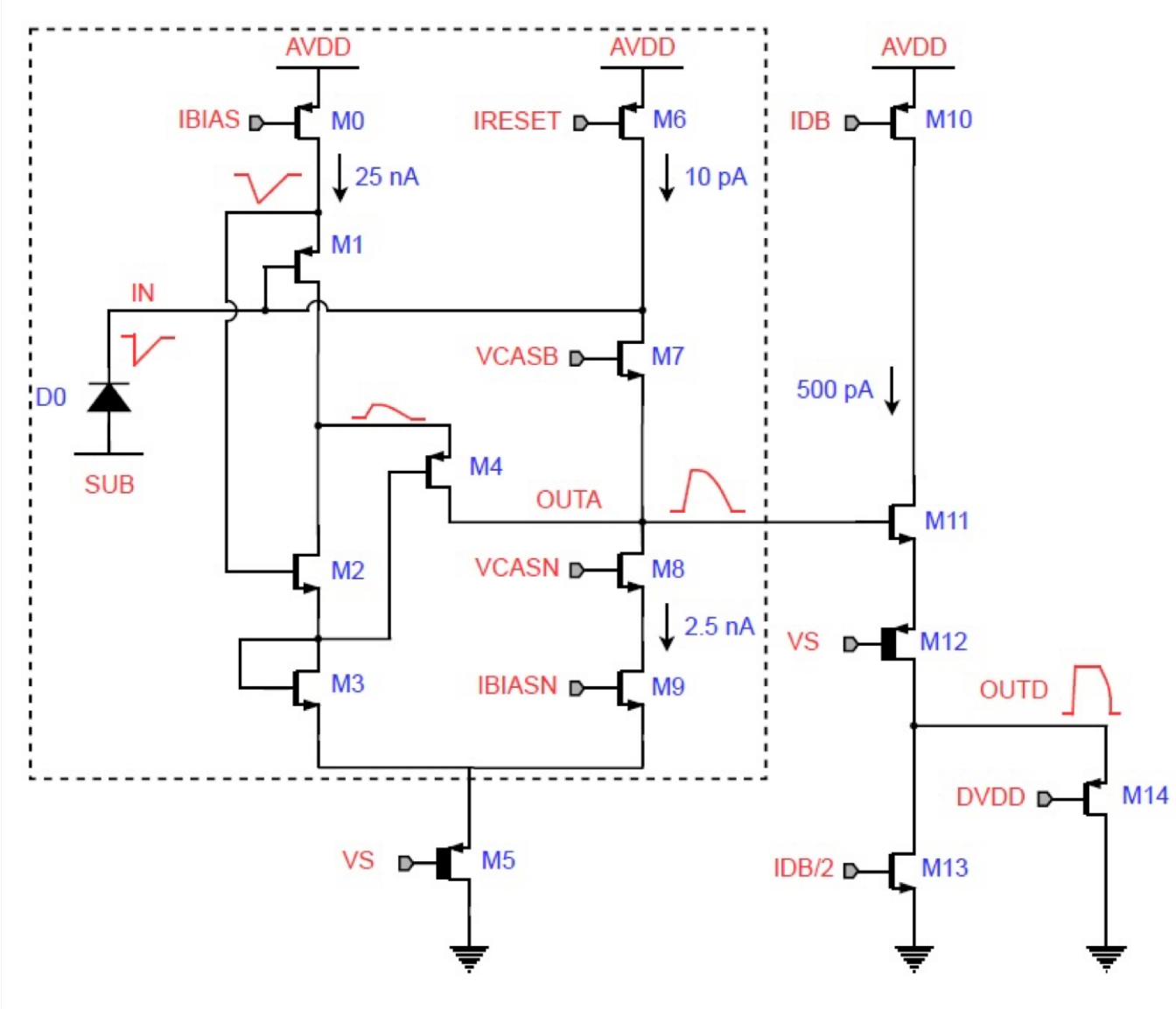}
        \label{fig:front_end}
    \end{subfigure}
    \hfill
    \begin{subfigure}[b]{0.45\textwidth}
        \centering
        \includegraphics[width=\textwidth]{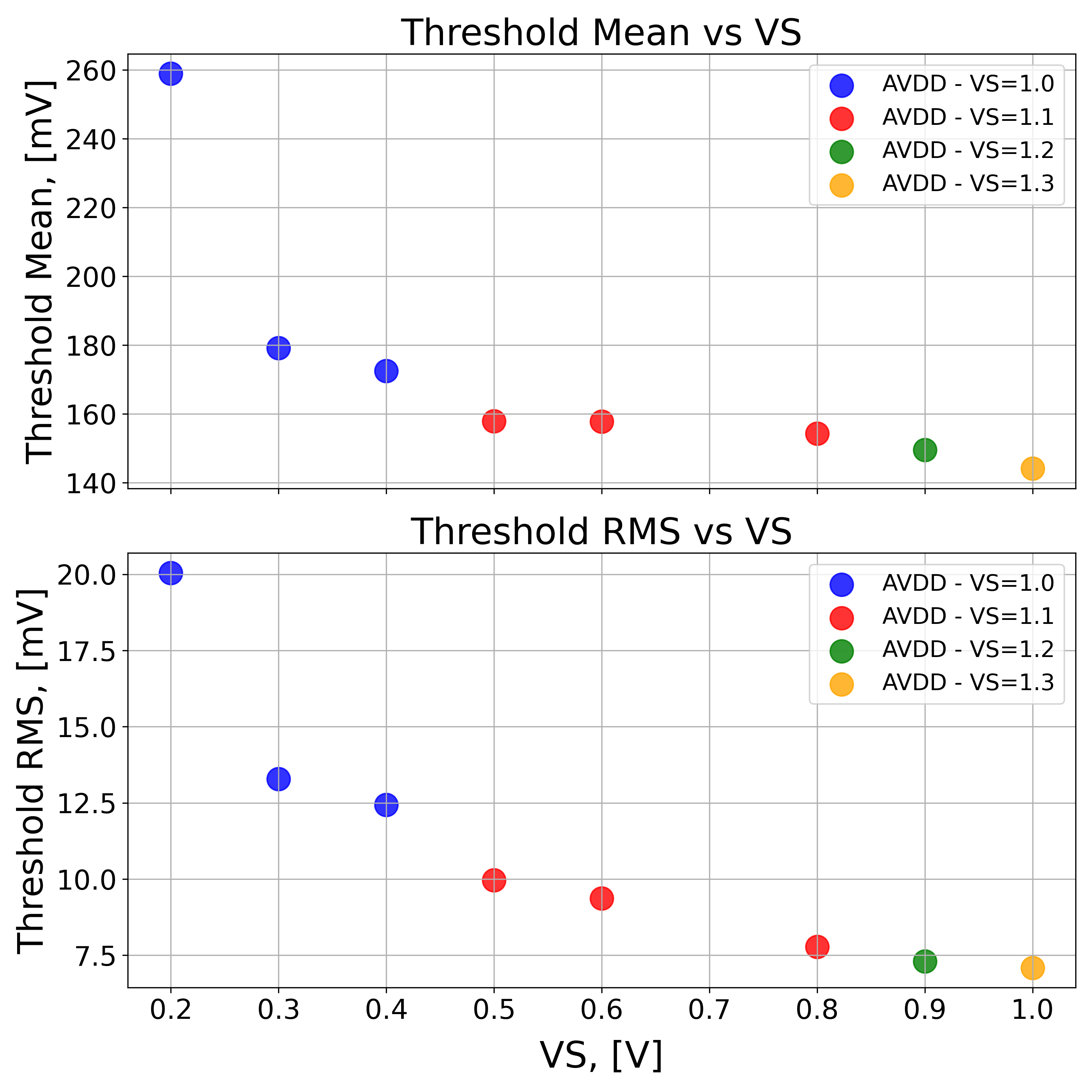}
        \label{fig:vs_analysis}
    \end{subfigure}
    \caption{Schematics of the MOST front-end electronics (left) and charge threshold and threshold spread versus VS (right).}
    \label{fig:frontend_vs}
\end{figure}
Reverse sensor bias is usually applied directly to the substrate. In the MOST sensor, however, reverse sensor bias is applied through the front-end electronics by adjusting the potential at its input. Increasing this input potential raises the reverse bias, with the option to increase the analog power supply if necessary. With this approach one  can connect the substrate directly to the analog circuit ground, simplifying the matrix design. The drawback is higher the power consumption when the analog power supply is increased. The reverse bias can be adjusted by varying $V_s$ in the MOST front-end, as illustrated in Figure~\ref{fig:frontend_vs} (left). Increasing $V_s$ forces the circuit to increase the potential at its input, and hence increases the sensor reverse bias.

Figure~\ref{fig:frontend_vs} (right) shows preliminary measurements with a trend confirming that increasing $V_s$ reduces charge threshold and threshold spread due to the decrease in sensor capacitance, as expected. The noise also decreases quickly to about 15 electrons or less, where it saturates at present. Further study of the operating point is in progress in an effort to reduce this saturation effect.

\section{Conclusion and Outlook}

The MOST sensor, a monolithic stitched sensor with timing capabilities, explores several design innovations with power gating, asynchronous readout, and an alternative sensor biasing scheme. The first results of the characterization have been presented here. A processing issue affected the overall yield, but the measurements nevertheless demonstrated the validity of the highly granular powering approach using a global power network and programmable power switches. The individual pixel yield above 99.98\% also confirms that this approach allows the in-pixel circuitry to be designed at the full density, very important for future designs.
First jitter and timing measurements show a functional asynchronous readout, but more work is needed to correctly evaluate its full timing performance. 
Preliminary measurements also illustrate the functionality of the front-end and its capability to reverse bias the sensor in an alternative way, allowing the substrate to be connected to the analog ground greatly simplifying the circuit design.

Future work will focus on further optimization of the chip operational parameters, detailed test beam campaigns to assess efficiency and spatial resolution, and expanded studies of its timing performance. 

\bibliographystyle{JHEP}
\bibliography{biblio.bib}

\providecommand{\href}[2]{#2}\begingroup\raggedright\begin{thebibliography}{1}

\bibitem{its3tdr}
{The ALICE Collaboration}, \emph{{Technical Design Report for the ALICE Inner Tracking System 3 - ITS3; A bent wafer-scale monolithic pixel detector}},  Tech. Rep. \href{https://cds.cern.ch/record/2890181}{CERN-LHCC-2024-003}, CERN, Geneva (2024).

\bibitem{apts_paper}
G.~Aglieri~Rinella, G.~Alocco, M.~Antonelli, R.~Baccomi, S.M.~Beole, M.B.~Blidaru et~al., \emph{{Characterisation of analogue Monolithic Active Pixel Sensor test structures implemented in a 65 nm CMOS imaging process}}, {\emph{preprint arXiv:2403.08952} (2024) }.

\bibitem{AGLIERIRINELLA2023168589}
G.~{Aglieri Rinella}, A.~Andronic, M.~Antonelli, M.~Aresti, R.~Baccomi, P.~Becht et~al., \emph{{Digital pixel test structures implemented in a 65 nm CMOS process}}, \href{https://doi.org/{https://doi.org/10.1016/j.nima.2023.168589}}{\emph{NIMA} {\bfseries 1056} (2023) 168589}.

\bibitem{SNOEYS201790}
W.~Snoeys et~al., \emph{{A process modification for CMOS monolithic active pixel sensors for enhanced depletion, timing performance and radiation tolerance}}, \href{https://doi.org/10.1016/j.nima.2017.07.046}{\emph{NIMA} {\bfseries 871} (2017) 90}.

\bibitem{leitao_development_2024}
P.V.~Leitao, \emph{Development of the {MOSAIX} chip for the {ALICE} {ITS3} upgrade},  in \emph{{TWEPP} 2024 {Topical} {Workshop} on {Electronics} for {Particle} {Physics}}, 2024, \href{{https://indi.to/jTDPS}}{{https://indi.to/jTDPS}}.

\bibitem{terlizzi_characterization_2024}
L.~Terlizzi, \emph{Characterization of {MOSS} for the {ALICE} {ITS3} for the {LHC} {Run} 4},  in \emph{Eleventh {International} {Workshop} on {Semiconductor} {Pixel} {Detectors} for {Particles} and {Imaging}}, 2024, \href{https://indico.in2p3.fr/event/32425/contributions/142817/}{https://indico.in2p3.fr/event/32425/contributions/142817/}.

\bibitem{piro_front-end_nodate}
F.~Piro, \emph{Front-end circuits for radiation-hard monolithic {CMOS} sensors targeting high-energy physics applications}, .

\bibitem{eberwein_yield_2024}
G.~Eberwein, \emph{{Yield Characterisation and Failure Analysis of the Monolithic Stitched Sensor\protect\nobreakspace{}MOSS for ALICE ITS3}},  in \emph{{TWEPP 2024 Topical Workshop on Electronics for Particle Physics}}, 2024, \href{https://indico.cern.ch/event/1381495/contributions/5988500/attachments/2938763/5162382/TWEPP2024\_vf\_pdf.pdf}{https://indico.cern.ch/event/1381495/contributions/5988500/attachments/2938763/5162382/TWEPP2024\_vf\_pdf.pdf}.

\end{thebibliography}\endgroup
\end{document}